\documentclass[prb,twocolumn,showpacs,superscriptaddress,floatfix]{revtex4}

\usepackage{graphicx}
\usepackage[centertags]{amsmath}

\newcommand{\cm}{\ensuremath{\,\mbox{cm}^{-1}}}
\newcommand{\K}{\ensuremath{\,\mbox{K}}}
\newcommand{\celsius}{\ensuremath{\,{}^\circ}\!C}

\begin{document}

\title{Magnetodielectric effect and phonon properties of compressively strained EuTiO$_{3}$
thin films deposited on LSAT}

\author{ S.~Kamba}\email{kamba@fzu.cz}
\affiliation{Institute of Physics ASCR, Na Slovance~2, 182 21 Prague~8, Czech Republic}
\author{V.~Goian}
\affiliation{Institute of Physics ASCR, Na Slovance~2, 182 21 Prague~8, Czech Republic}
\author{M. Orlita}
\affiliation{Laboratoire National des Champs Magn\'{e}tiques Intenses, CNRS-UJF-UPS-INSA,
25, avenue des Martyrs, 38042 Grenoble, France}
\author{D.~Nuzhnyy}
\affiliation{Institute of Physics ASCR, Na Slovance~2, 182 21 Prague~8, Czech Republic}
\author{J.H. Lee} \affiliation{Department of Materials Science and
Engineering, Cornell University, Ithaca, New York, 14853-1501, USA}
\author{D.G.~Schlom} \affiliation{Department of Materials Science and
Engineering, Cornell University, Ithaca, New York, 14853-1501, USA}
\affiliation{Kavli Institute at Cornell for Nanoscale Science, Ithaca, New York 14853, USA}
\author{K.Z.~Rushchanskii}
\affiliation{Peter Gr\"{u}nberg Institut, Quanten-Theorie der Materialien,
Forschungszentrum J\"{u}lich GmbH, 52425 Juelich and JARA FIT, Germany}
\author{M. Le\v{z}ai\'{c}}
\affiliation{Peter Gr\"{u}nberg Institut, Quanten-Theorie der Materialien,
Forschungszentrum J\"{u}lich GmbH, 52425 Juelich and JARA FIT, Germany}
\author{T. Birol}
\affiliation{School of Applied and Engineering Physics, Cornell University, Ithaca, New
York, 14853, USA}
\author{C.J. Fennie}
\affiliation{School of Applied and Engineering Physics, Cornell University, Ithaca, New
York, 14853, USA}
\author{ P. Gemeiner }
\affiliation{Laboratoire Structures, Propri\'{e}t\'{e}s et Mod\'{e}lisation des Solides,
UMR8580 CNRS-Ecole Centrale Paris, 92295 Ch\^{a}tenay-Malabry Cedex, France}
\author{B. Dkhil}
\affiliation{Laboratoire Structures, Propri\'{e}t\'{e}s et Mod\'{e}lisation des Solides,
UMR8580 CNRS-Ecole Centrale Paris, 92295 Ch\^{a}tenay-Malabry Cedex, France}
\author{V. Bovtun}
\affiliation{Institute of Physics ASCR, Na Slovance~2, 182 21 Prague~8, Czech Republic}
\author{M. Kempa}
\affiliation{Institute of Physics ASCR, Na Slovance~2, 182 21
Prague~8, Czech Republic}
\author{J.~Hlinka}
\affiliation{Institute of Physics ASCR, Na Slovance~2, 182 21 Prague~8, Czech Republic}
\author{J.~Petzelt}
\affiliation{Institute of Physics ASCR, Na Slovance~2, 182 21
Prague~8, Czech Republic}

\date{\today}


\begin{abstract}

Compressively strained epitaxial (001) EuTiO$_{3}$ thin films of tetragonal symmetry have
been deposited on (001) (LaAlO$_{3}$)$_{0.29}$-(SrAl$_{1/2}$Ta$_{1/2}$O$_{3}$)$_{0.71}$
(LSAT) substrates by reactive molecular-beam epitaxy. Enhancement of the N\'{e}el
temperature by 1\,K\, with 0.9\% compressive strain was revealed. The polar phonons of
the films have been investigated as a function of temperature and magnetic field by means
of infrared reflectance spectroscopy. All three in-plane polarized infrared active
phonons show strongly stiffened frequencies compared to bulk EuTiO$_{3}$ in accordance
with first principles calculations. The phonon frequencies exhibit gradual softening on
cooling leading to an increase in static permittivity. A new polar phonon with frequency
near the TO1 soft mode was detected below 150\K. The new mode coupled with the TO1 mode
was assigned as the optical phonon from the Brillouin zone edge, which is activated in
infrared spectra due to an antiferrodistortive phase transition and due to simultaneous
presence of polar and/or magnetic nanoclusters. In the antiferromagnetic phase we have
observed a remarkable softening of the lowest-frequency polar phonon under an applied
magnetic field, which qualitatively agrees with first principles calculations. This
demonstrates the strong spin-phonon coupling in EuTiO$_{3}$, which is responsible for the
pronounced dependence of its static permittivity on magnetic field in the
antiferromagnetic phase.

\end{abstract}

\maketitle

\section{Introduction}

Multiferroic compounds in which magnetic and electric orders coexist are intensively
studied due to their high potential in magnetoelectric devices as well as due to their
rich and fascinating physics related to the magnetoelectric effect. Unfortunately, it
appears that there are only few multiferroics in the nature and their absolute majority
exhibits multiferroic properties deeply below room temperature.\cite{wang09} Moreover,
most multiferroics exhibit antiferromagnetic (AFM) order and therefore only weak
magnetoelectric coupling.\cite{wang09} For this reason, there is an intensive search for
new materials with ferromagnetic and ferroelectric order, where the magnetoelectric
coupling could be high.

Fennie and Rabe\cite{fennie06} suggested a new route for preparation of multiferroics
with a strong magnetoelectric coupling. They proposed to use a biaxial strain in the thin
films for induction of the ferroelectric and ferromagnetic state in materials which are
in the bulk form paraelectric and AFM. Basic condition for such a material is the strong
spin-phonon coupling. Fennie and Rabe\cite{fennie06} proposed, based on their first
principles calculations, to use EuTiO$_{3}$ for such purpose. Recently, Lee et
al.\cite{lee10} actually confirmed the theoretical prediction and revealed ferroelectric
and ferromagnetic order in the tensile strained EuTiO$_{3}$ thin films deposited on
DyScO$_{3}$ substrates.

Bulk EuTiO$_{3}$ is an antiferromagnet with G-type AFM order below $T_{N}$=
5.3\,K.\cite{guire66} Temperature dependence of its permittivity $\varepsilon$' exhibits
quantum paraelectric behavior, i.e. $\varepsilon$' increases on cooling and saturates at
low temperatures due to quantum fluctuations, which inhibit the creation of long-range
ferroelectric order. Recently, it was theoretically predicted that EuTiO$_{3}$ has
antiferrodistorted ground state.\cite{rushchanskii10,rushchanskii11} Experimental studies
show that at room temperature the crystal structure is cubic $Pm\overline{3}m$
perovskite,\cite{brous53} which transforms to tetragonal $I4/mcm$ structure near
280\K.\cite{bussmann-holder11,allieta11,goian12a} $\varepsilon$'(T) exhibits a sharp drop
below $T_{N}$ due to a strong spin-phonon coupling.\cite{katsufuji01} A large interaction
of the magnetic moment with the crystal lattice was also manifested by 7\% increase of
$\varepsilon$' with magnetic field, observed at 2\K.\cite{katsufuji01} Linear
magnetoelectric coupling is forbidden by the symmetry of EuTiO$_{3}$. In spite of this, a
strong bi-quadratic magnetoelectric coupling was recently reported below
$T_{N}$.\cite{shvartsman10}

Low-temperature infrared (IR) reflectivity studies of EuTiO$_{3}$
ceramics reveal three polar phonons typical for cubic
perovskites.\cite{kamba07,goian09}
The frequency of the lowest frequency (TO1) phonon decreases (the mode softens) on
cooling, which fully explains the increase of static permittivity on lowering temperature
via the Lyddane-Sachs-Teller (LST) relation. Below $\approx$ 100\K\, the soft TO1 mode
frequency deviates from the classical Cochran law and finally levels off below
30\K.\cite{kamba07} Such temperature dependence follows the Barrett formula, which
describes the saturation of $\varepsilon$'(T) at low temperatures due to quantum
fluctuations. Based on the plasma frequencies of the polar phonons, the TO1 mode was
assigned to Slater mode, i.e. Ti cations vibration against the oxygen octahedra, and the
TO2 phonon (near 150\cm) was assigned to Last mode, i.e. vibration of Eu cations against
the TiO$_{6}$ octahedra.\cite{goian09} Temperature dependence of the IR spectra indicated
coupling (mixing of their eigenvectors) of both the modes.\cite{goian09} Moreover, a
strong spin coupling with the soft mode (SM)is expected microscopically due to a
superexchange between Eu$^{2+}$ 4$f$ spins via the 3$d$ states of the Ti$^{4+}$
ions.\cite{akamatsu11}

Here we report a study of the epitaxial compressively strained EuTiO$_{3}$ thin films
deposited on (001) (LaAlO$_{3}$)$_{0.29}$-(SrAl$_{1/2}$Ta$_{1/2}$O$_{3}$)$_{0.71}$ (LSAT)
substrate. Lee et al.\cite{lee10} have shown that the maximum expected compressive strain
of -0.9\% in this case is not sufficient for induction of the ferroelectric and
ferromagnetic order. Nevertheless, we will show here that the phonon properties of
EuTiO$_{3}$/LSAT are strongly influenced: The in-plane soft mode frequency is remarkably
stiffened in comparison to the bulk EuTiO$_{3}$ ceramics and moreover its IR reflection
band is more than five times narrower. It enabled us a detailed investigation of the
magnetic field dependence of this phonon. We will show that shift of the lowest-frequency
optical phonon with magnetic field is mainly responsible for observed magnetodielectric
effect in EuTiO$_{3}$. Moreover, we will report about an antiferrodistortive (AFD) phase
transition appearing in this film near 150\K.

\section{Experimental}

Films of two thicknesses (22 and 42 nm) were deposited by reactive molecular-beam epitaxy
on the (001) LSAT substrate. Details of the deposition were described
elsewhere.\cite{lee09} The 42\,nm film was slightly relaxed, but the 22\,nm film was
fully (-0.9\%) compressively strained, its XRD showing the same in-plane lattice
parameter (3.870 \AA) as the substrate. The substrates of 10x10x1 mm size were provided
by R. Uecker and his colleagues from the Institute of Crystal Growth, Berlin, Germany.

The unpolarized IR reflectance spectra were taken using a Bruker
IFS 113v FTIR spectrometer at temperatures from 1.8 to 300\K\,
with the resolution of 2\cm. We checked also the polarized spectra
at 300 and 10 K, but no in-plane anisotropy was observed,
confirming that the substrate plate as well as the thin films are
macroscopically optically isotropic in the (001) plane down to
10\K. An Optistat CF cryostat (Oxford Instruments) was used for
cooling the samples without magnetic field down to 5\K. The
investigated spectral range (up to 650\cm) was limited by the
transparency region of the polyethylene windows of the cryostat. A
helium-cooled Si bolometer operating at 1.6\K\, was used as a
detector. Custom-made superconducting magnetic cryostat was used
for IR studies at various magnetic fields up to 13\,T at
temperatures between 1.8 and 4.2\K.

Each of the reflectance spectra was evaluated as a two-layer
optical system.\cite{zelezny98} At first, the bare substrate
reflectivity was measured as a function of temperature and
carefully fitted using the generalized factorized damped harmonic
oscillator model\cite{gervais83}
\begin{equation}\label{eps4p}
\varepsilon^{*}(\omega)=\varepsilon_{\infty}\prod_{j=1}^n\frac{\omega^{2}_{LOj}-\omega^{2}+i\omega\gamma_{LOj}}{\omega^{2}_{TOj}-\omega^{2}+i\omega\gamma_{TOj}}
\end{equation}
where $\omega_{TOj}$ and $\omega_{LOj}$ denote the transverse and longitudinal frequency
of the j-th polar phonon, respectively, and $\gamma$$_{TOj}$ and $\gamma$$_{LOj}$ denote
their corresponding damping constants. $\varepsilon$$^{*}$($\omega$) is related to the
reflectivity R($\omega$) of the bulk substrate by
\begin{equation}\label{refl}
R(\omega)=\left|\frac{\sqrt{\varepsilon^{*}(\omega)}-1}{\sqrt{\varepsilon^{*}(\omega)}+1}\right|^2.
\end{equation}
The high-frequency permittivity $\varepsilon_{\infty}$ = 5.88 resulting from the
electronic absorption processes was obtained from the room-temperature
frequency-independent reflectivity tails above the phonon frequencies
 and was assumed to be temperature independent.

When analyzing the reflectance of the substrate together with the
film, we kept the earlier fitted bare substrate parameters fixed
at each temperature and only the dielectric function of the film
was adjusted. For this purpose, we preferentially used a classical
three-parameter damped oscillator model\cite{gervais83}
\begin{equation}
\label{eps3p}
 \varepsilon^*(\omega)
 = \varepsilon_{\infty} + \sum_{j=1}^{n}
\frac{\Delta\varepsilon_{j}\omega_{TOj}^{2}} {\omega_{TOj}^{2} -
\omega^2+\textrm{i}\omega\gamma_{TOj}} \, ,
\end{equation}
since it uses less fitting parameters. This approach is well justified, because the
damping of LO phonons of the film do not influence appreciably the reflectance spectra.
Only the lineshape of the lowest frequency doublet was not well reproduced and therefore
a coupled damped harmonic model was applied at low temperatures (see below.)

The IR reflectance measurements were performed on samples with a 1\,mm thick substrates
which was thick enough to avoid the parasitic IR signal from the multiple reflection in
the substrate. The same samples were afterwards polished to 248$\pm 1 \mu$m to reach the
optimal MW resonance near 15\,GHz where the samples were measured as the composite
TE$_{01\delta}$ dielectric resonators in the shielding cavity\cite{bovtun11} using their
dielectric resonance. Precisely the same thicknesses of bare substrates and substrates
with the films are needed for an accurate evaluation of the in-plane complex permittivity
of the films. Each sample was measured twice, in the Sigma System M18 temperature chamber
(100-370\K) and in the He-cooled Janis close-cycle cryostat (10-370\K).

High resolution X-ray diffraction (XRD) measurements were
performed on a two-axis diffractometer in a Bragg-Brentano
geometry (focalization circle with a diameter of 50\,cm).
Cu-K$\alpha$ (wavelength $\lambda$=1.54056$\AA$) radiation was
emitted from a 18\,kW rotating anode generator. The out of plane
lattice constant of the film as well as the substrate were
determined with accuracy of $\pm$0.0003 $\AA$. Cryostat was
operating between 10\K\, and 300\K\, with an accuracy better than
0.5\K.

Magnetic properties were measured using a SQUID magnetometer
(Quantum Design) from 2 to 30\K. Sample with the 42\,nm thin film
was cut to four pieces of 3.5x4.5x0.248 mm$^{3}$ size and all four
stacked samples were measured simultaneously.

\section{Theoretical considerations}

In parallel with the experimental studies, we also investigated
the EuTiO$_{3}$ films using a first-principles density-functional
calculations within the spin-polarized gradient approximation
(GGA). In particular, we have studied the effect of the strain and
magnetic field on phonon frequencies in the bulk and thin films.
Our initial guess for the low temperature structure of EuTiO$_{3}$
is analogous to that of the bulk SrTiO$_3$ ($I4/mcm$ space group).
In the course of this work, this structure was confirmed
experimentally\cite{goian12a} for the bulk EuTiO$_{3}$ and
according to our calculations, it is also a plausible structure
for the thin films under the compressive strains.

 For the electronic structure
calculations and structural relaxations we used projector augmented-wave potentials as
implemented in the Vienna \textit{Ab initio} Simulation Package (VASP). To account for
the strong electron correlation effects on the $f$-shell of Eu atoms, we used the DFT+U
scheme in the Dudarev's approach with an on-site Hubbard U and Hund's exchange J$_{H}$.
For the bulk EuTiO$_3$ of space group $Pm\bar{3}m$, U=5.7 eV is used. In order to correct
for the effect of oxygen octahedron rotations on the magnetism, the Hubbard parameter is
increased to U=6.2 eV for EuTiO$_3$ in the space group $I4/mcm$.\cite{birolETO} Hund's
exchange is kept fixed at J$_H$=1.0 eV for all calculations. Details of the calculations
can be found in Refs. \cite{rushchanskii10,rushchanskii11}.

For the sake of comparison with the experiment, we have calculated frequencies of the
polar phonon modes in the bulk cubic and tetragonal EuTiO$_3$ as well as in tetragonal
lattices with equatorial lattice parameters fixed to those of the SrTiO$_3$ and LSAT
substrates. In order to estimate the frequency changes under strong bias magnetic field,
the calculations were carried out both for the AFM order (anticipated ground state) and
for ferromagnetic (FM) order (i.e. for the expected induced magnetic order under magnetic
field). Resulting mode frequencies are listed in Table~\ref{TableFreqs62}.

\begin{widetext}
\begin{table}[h]
\centering{}
\begin{tabular}{||cc||cc|cc||cc|cc||cc|cc||c|c||}
\hline
\multicolumn{2}{||c||}{Bulk EuTiO$_3$} & \multicolumn{4}{c||}{Bulk EuTiO$_3$} & \multicolumn{4}{c||}{EuTiO$_3$/SrTiO$_3$}&\multicolumn{4}{c||}{EuTiO$_3$/LSAT}&\multicolumn{2}{c||}{EuTiO$_3$/LSAT}\\
\multicolumn{2}{||c||}{$Pm\bar{3}m$} &  \multicolumn{4}{c||}{$I4/mcm$} & \multicolumn{4}{c||}{$I4/mcm$}&\multicolumn{4}{c||}{$I4/mcm$}&\multicolumn{2}{c||}{Experiment at 1.9\K}\\
\hline\hline
\multicolumn{2}{||c||}{$F_{1u}$}&\multicolumn{2}{c}{$A_{2u}$}&\multicolumn{2}{|c||}{$E_u$}&\multicolumn{2}{c}{$A_{2u}$}&\multicolumn{2}{|c||}{$E_u$}&\multicolumn{2}{c}{$A_{2u}$}&\multicolumn{2}{|c||}{$E_u$}&\multicolumn{1}{c|}{$E_u$}&\multicolumn{1}{c||}{$E_u$}\\
FM & AFM & FM&AFM& FM&AFM& FM&AFM& FM&AFM& FM&AFM& FM&AFM& FM&AFM \\
\hline
70&77& 125&133&    107&113&    132&138&    77&86&      113&124&    121&125&    104.0, ? & 105.5, 115\\
162&165&    158&159&    156&158&    159&161&    147&148&    160&161&    164 & 166 & ?&162 \\
 & &  & &    251&251&    & &    246&246&    & &    255&255&    & \\
&&     &&          422&422&    &&          414&414&    &&          428&428&   & \\
549&550& 533&534&          526&527&536    &537&          519&519& 531   &532 &          532&532&  ?&  551\\
\hline
\end{tabular}
\caption{Polar phonon frequencies in \cm\, of EuTiO$_3$ from the
first principles for the bulk and for films on two different
substrates, SrTiO$_3$ (i.e. no strain) and LSAT (compressive
strain -0.9\%). Experimental results are given in the last column
for comparison. Only the lowest-frequency phonon was studied as
dependent on magnetic field. The calculations are valid at 0\K. If
the oxygen octahedron rotations \textit{are not} taken into
account, the structure of the bulk EuTiO$_3$ is $Pm\bar{3}m$. If
the oxygen octahedron rotations \textit{are} taken into account,
the space group is ($I4/mcm$), the same both in bulk and thin
films. The in-plane polarized polar modes have the $E_u$ symmetry,
whereas the out-of-plane modes have the $A_{2u}$ symmetry.}
\label{TableFreqs62}
\end{table}
\end{widetext}

\section{Results}

\subsection{Magnetic properties}
\begin{figure}
  \begin{center}
    \includegraphics[width=80mm]{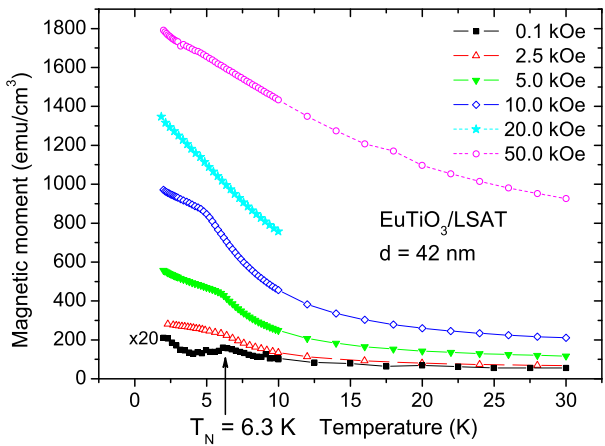}
  \end{center}
    \caption{(Color online) Temperature dependence of the magnetization of 42\,nm thick film at
    different external magnetic fields.}
    \label{Fig1}
\end{figure}

Magnetization curves taken at various external magnetic fields (see Fig.~\ref{Fig1})
reveal T$_{N}$=6.3\K. It means that the -0.9\% compressive strain in the film enhances
T$_{N}$ by 1.0\K\, because the coupling of spins localized at the 4\textit{f} levels of
Eu cations increases with the reducing lattice parameter in the compressively strained
films. Note that the +1.1\% tensile strain in EuTiO$_{3}$/DyScO$_{3}$ induces the
ferromagnetic order below 4.2\K.\cite{lee10} On the other hand, it was reported that the
as-deposited thin EuTiO$_{2.86}$ films on SrTiO$_{3}$ (this substrate has exactly the
same lattice constant 3.905\,\AA\, as the bulk EuTiO$_{3}$) exhibit an elongated
\textit{c}-axis and a ferromagnetic order below 5\,\K,\cite{Kugimiya07,Fujita09} while
the same films post annealed at 1000\,\celsius\, in reducing atmosphere exhibit AFM order
and relaxed out-of-plane lattice strain.\cite{Fujita09}

\subsection{Temperature dependence of the phonon spectra}
\begin{figure}
  \begin{center}
    \includegraphics[width=80mm]{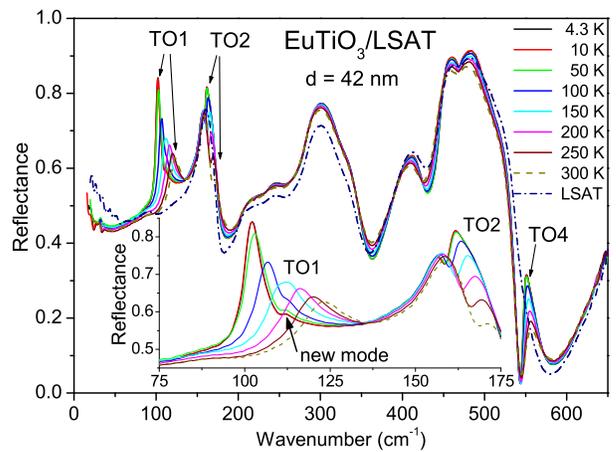}
  \end{center}
    \caption{(Color online) Temperature dependence of the IR reflectance of the EuTiO$_{3}$ 42\,nm film
    deposited on the LSAT substrate. Room-temperature reflectivity of pure LSAT substrate is also plotted
    for comparison (dashed-dotted line). Phonons from the thin film are marked. Inset shows the enlarged low-frequency part of
    the reflectance spectra, where the softening of TO1 and TO2
    phonons is clearly seen.}
    \label{Fig2}
\end{figure}

\begin{figure}
  \begin{center}
    \includegraphics[width=80mm]{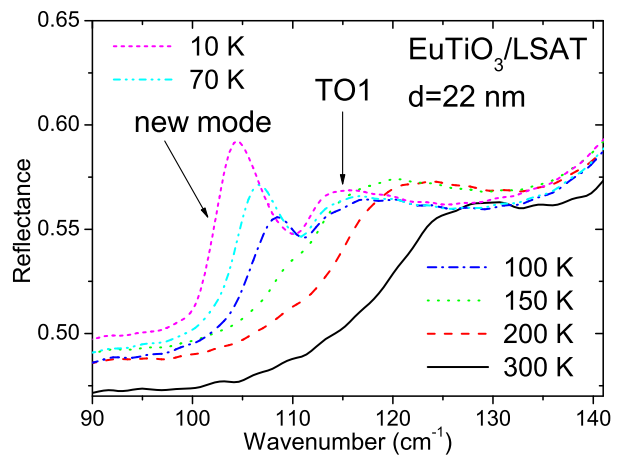}
  \end{center}
    \caption{(Color online) Temperature dependence of the low-frequency IR reflectance of fully strained EuTiO$_{3}$
    22\,nm film deposited on a LSAT substrate. Remarkable TO1 phonon softening on cooling and activation of a new mode below 150\K\,
    is clearly seen.}
    \label{Fig3}
\end{figure}

IR reflectance of the EuTiO$_{3}$ 42 nm film plotted at various
temperatures is shown in Fig.~\ref{Fig2}. Comparison of the
spectra with those of bare LSAT substrate allows us to distinguish
the three TO phonons of the film (marked by the arrows). The
highest frequency phonon is marked TO4 (the TO3 phonon is silent).
The two lowest-frequency TO phonons show up softening on cooling,
which is better seen in the inset of Fig.~\ref{Fig2}. The three
modes correspond well the three polar modes of the cubic
EuTiO$_{3}$, as shown in the Table\,I. Moreover, a new weak mode
appears near 112\cm\, below 100\K. Since the 42\,nm film was not
fully strained, we have measured also the IR reflectance spectra
from a fully (and uniformly) strained 22\,nm film. Due to the
smaller film thickness, the intensity of the TO1 reflection band
is lower (see Fig.~\ref{Fig3}), but one can clearly see the mode
splitting already below 150\K. Note that in this case the new mode
is activated below the TO1 phonon frequency, while in the thicker
film the new mode has higher frequency than the TO1 mode. Origin
of the new mode will be discussed in section V.

Shape of the IR reflectance bands in Fig.~\ref{Fig3} is rather
unusual and the splitting of the TO1 mode cannot be fitted well
just with a sum of independent oscillators (Eq.~\ref{eps3p}).
Therefore, the coupled oscillator formula was used for the fit of
the split modes (their bare parameters are marked with subscripts
$a$ and $b$) and the rest of modes were fitted with
Eq.~\ref{eps3p}. The coupled oscillator formula has the form
\cite{petzelt87}
\begin{widetext}
\begin{equation}
\label{coupled}
 \varepsilon^*(\omega)
 = \varepsilon_{\infty} +
\frac{\Delta\varepsilon_{1a}\omega_{TO1a}^{2}(\omega_{TO1b}^{2}-\omega^{2}+
\textrm{i}\omega\gamma_{TO1b})
+\Delta\varepsilon_{1b}\omega_{TO1b}^{2}(\omega_{TO1a}^{2}-\omega^{2}+
\textrm{i}\omega\gamma_{TO1a})-2i\omega\Gamma\sqrt{\Delta\varepsilon_{1a}\omega_{TO1a}^{2}\Delta\varepsilon_{1b}\omega_{TO1b}^{2}}}
{(\omega_{TO1a}^{2} - \omega^2+\textrm{i}\omega\gamma_{TO1a})(\omega_{TO1b}^{2} -
\omega^2+\textrm{i}\omega\gamma_{TO1b})+\omega^{2}\Gamma^{2}} \,
\end{equation}
\end{widetext}
using the imaginary coupling constant $\Gamma$ model. The rest of symbols have the same
meaning as in Eq.~\ref{eps3p}.

\begin{figure}
  \begin{center}
    \includegraphics[width=83mm]{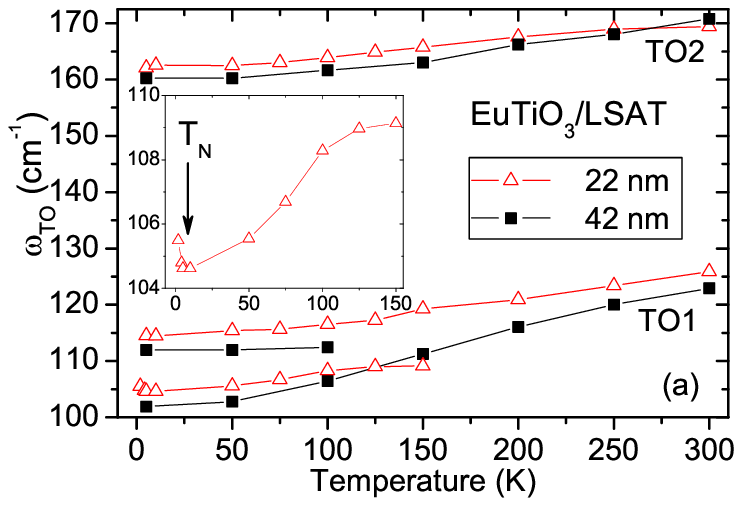}
    \includegraphics[width=83mm]{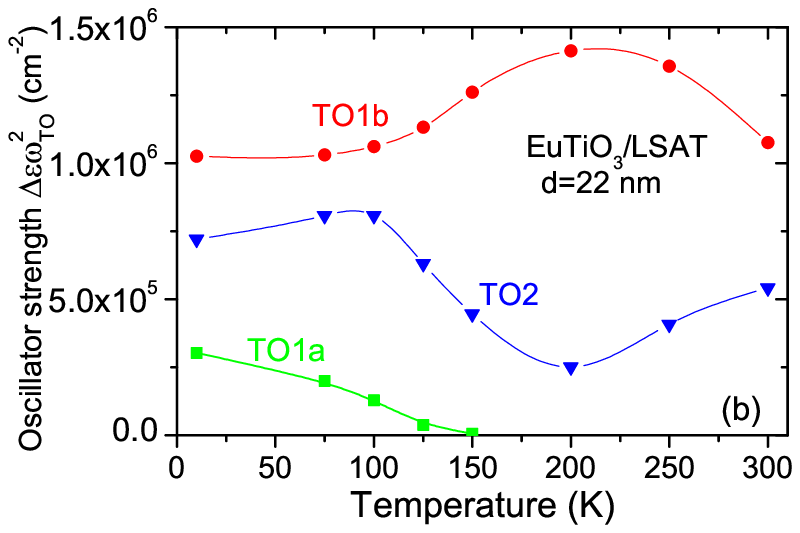}
  \end{center}
    \caption{(Color online) a) Temperature dependence of the phonon frequencies for the EuTiO$_{3}$ films
    of two thicknesses 42 and 22\,nm below 170\cm.  Stiffening of the TO1 and TO2 phonon
    frequencies in the more strained 22 nm film is seen. The new mode in the 22\,nm film activates by 50\K\, higher than
    in the thicker film. The inset shows the temperature dependence of the
    new mode frequency in expanded scale down to 1.9\K. Mode hardening below the N\'{e}el temperature is clearly seen.
    b) Temperature dependence of the oscillator strengths of all the modes seen below 200\cm. One can see that the
    new TO1a mode gains strength from the TO1b mode.}
    \label{Fig4}
\end{figure}

\begin{figure}
  \begin{center}
    \includegraphics[width=85mm]{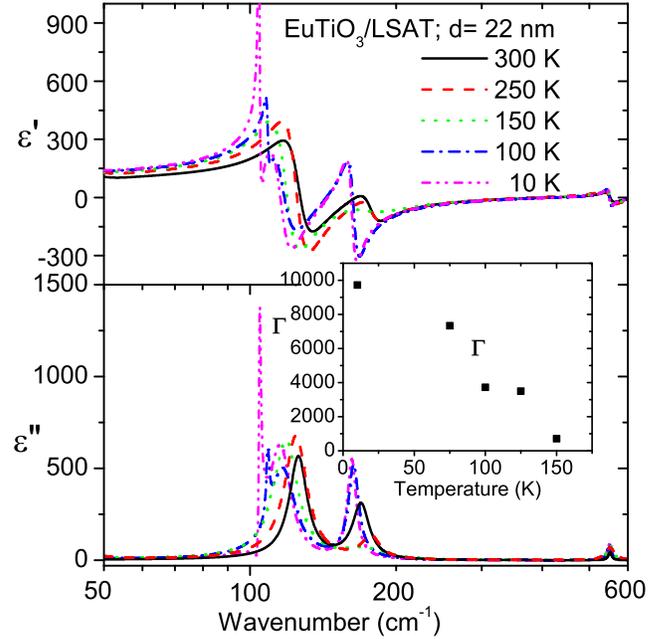}
  \end{center}
    \caption{(Color online) Complex dielectric function spectra of 22\,nm EuTiO$_{3}$ film evaluated at various
    temperatures from the reflectance spectra in Fig.~\ref{Fig3}. The inset shows
    the temperature dependence of the imaginary coupling constant between two lowest-frequency modes.}
    \label{Fig5}
\end{figure}

Fig.~\ref{Fig4}a shows the temperature evolution of the polar phonon frequencies below
170\cm\, obtained from the reflectance fits of both films. At 10\K\, the TO phonon
frequencies are 83, 151 and 533\cm\, for the bulk ceramics,\cite{goian09} while our
22\,nm film shows phonons at 115, 162, 551 \cm\, and the additional new mode at 105 \cm.
There is a strong phonon frequency stiffening in the films due to the compressive strain.
Note also that the phonon frequencies are higher in the 22\,nm film than in the 42\,nm
one (see Fig.~\ref{Fig4}a), obviously due to the higher strain in the former case.

Complex permittivity spectra evaluated from the IR reflectance are
shown in Fig.~\ref{Fig5}. Stiffening of all in-plane polar phonons
is responsible for a lower in-plane permittivity $\varepsilon$' in
the films compared with the bulk sample. At 10\K\, the static
permittivity reaches the value of 146 compared to 405 in the
single crystal.\cite{katsufuji01} Here it should be stressed that
compressive strain reduces the in-plane permittivity, but the
out-of-plane permittivity should be enhanced, because the TO1
phonon component polarized perpendicularly to the film plane
should be softer. Unfortunately, we cannot see this phonon in our
near-normal reflectivity geometry.

In our fit we used $\Delta\varepsilon_{1a}$=0, i.e. we suppose
that the bare lowest-frequency mode has zero dielectric strength.
The imaginary coupling does not shift the phonon frequencies, it
merely deforms the spectral line shape and transfers the
oscillator strength $\Delta\varepsilon\omega_{TOj}^{2}$ from TO1b
to the new TO1a mode (see Fig.~\ref{Fig4}b). Dramatic increase of
the coupling constant $\Gamma$ is seen on cooling below 150\K\, in
inset of Fig.~\ref{Fig5}. This causes the increase of its strength
on cooling. Also the temperature dependences of the TO1b and TO2
mode oscillator strengths are remarkable. They give evidence on
mixing of eigenvectors of both the modes.

\subsection{Microwave dielectric measurements}

\begin{figure}
  \begin{center}
    \includegraphics[width=80mm]{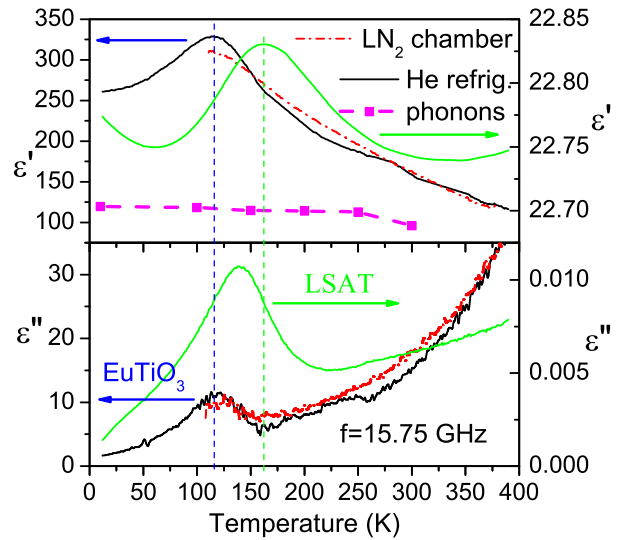}
  \end{center}
    \caption{(Color online) Temperature dependence of the complex permittivity in  the 22\,nm\, EuTiO$_{3}$ film (left
    scale) and the LSAT substrate (right scale) measured at 15.75\,GHz. The thin film MW data
    were obtained using two low-temperature devices (LN$_{2}$ chamber and He refrigerator). Static permittivity calculated from
    the phonon contributions (solid dots) is shown for comparison.}
    \label{Fig7}
\end{figure}

The non-monotonous dependence of the oscillator strengths of TO1b
and TO2 modes (see Fig.~\ref{Fig4}b) could be related to an AFD
phase transition involving anti-phase tilts of the oxygen
octahedra ($a^{0}a^{0}c^{-}$ in Glazer notation). This phase
transition occurs in the bulk EuTiO$_{3}$ near room
temperature\cite{bussmann-holder11,goian12a}, but in the strained
films it can be expected at lower temperatures (e.g. near 180\K\,
in the 1\% tensile strained EuTiO$_{3}$/DyScO$_{3}$)\cite{lee10}.
Also the TO1a -TO1b coupling could be related to this transition,
if the distorted phase would be ferroelectric. Therefore, we have
investigated the temperature dependence of the in-plane MW
permittivity of the film and of the substrate itself. Complex
permittivity of the film measured near 15 GHz shows a dielectric
anomaly near 120\K\,(see Fig.~\ref{Fig7}) reminding a
ferroelectric phase transition. Nevertheless, the second harmonic
generation (SHG) measurements of the EuTiO$_{3}$/LSAT film did not
reveal any acentric phase down to 2\K.\cite{lee10} Moreover, from
theory we know that the spontaneous polarization (and the main
dielectric anomaly) should occur in the [001] direction in the
compressively strained films,\cite{fennie06,schlom07}, whereas we
measure only the in-plane dielectric response. It is suspicious
that the low temperature MW $\varepsilon$' is higher than the sum
of phonon contributions to $\varepsilon$' (see Fig.~\ref{Fig7})
and that the maximum of $\varepsilon$' occurs 30\K\, below the
temperature where the new mode activates in the IR spectra. Since
we do not see any phonon anomaly in the IR spectra explaining the
dielectric anomaly near 120\K, it seems that the MW dielectric
anomaly has its origin in some dielectric dispersion below the
phonon frequencies, most probably due to diffusion of charged
defects (e.g. oxygen vacancies). Such defects play the main role
in the conductivity of the films, which deteriorates the
dielectric studies of EuTiO$_{3}$/LSAT in the kHz region.

\subsection{Magnetodielectric effect and tuning of the phonon frequency by magnetic field}

Bulk EuTiO$_{3}$ has a high static $\varepsilon'\simeq$\,400 and low SM frequency.
Therefore it shows a high (7\%) tunability of $\varepsilon$' by the magnetic field
\textit{B},\cite{katsufuji01} and an unusually high 7\cm\, change of the SM TO1 frequency
with \textit{B} is expected theoretically.\cite{fennie06} Such a shift was not confirmed
in our IR spectra of the bulk ceramics, presumable because the TO1 reflection band is too
broad (almost 100\cm\,). \cite{kamba07} However, the sharper TO1 reflectance band in
compressively strained EuTiO$_{3}$ films (see Figs.~\ref{Fig2} and ~\ref{Fig3}) appears
to be more promising for such observation, although the expected frequency shift (see
Table~\ref{TableFreqs62}) is somewhat lower ($\approx$ 4\cm).

\begin{figure}
  \begin{center}
    \includegraphics[width=80mm]{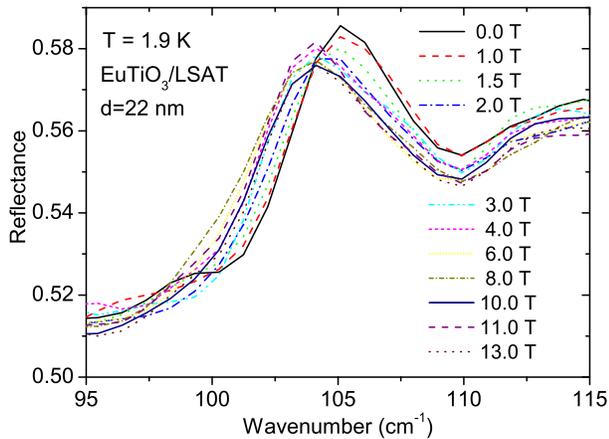}
  \end{center}
    \caption{(Color online) IR reflectance spectra of the 22\,nm thin film at 1.9\,K taken at various magnetic fields. The shift of
    phonon frequency is clearly seen.}
    \label{Fig9}
\end{figure}

\begin{figure}
  \begin{center}
    \includegraphics[width=70mm]{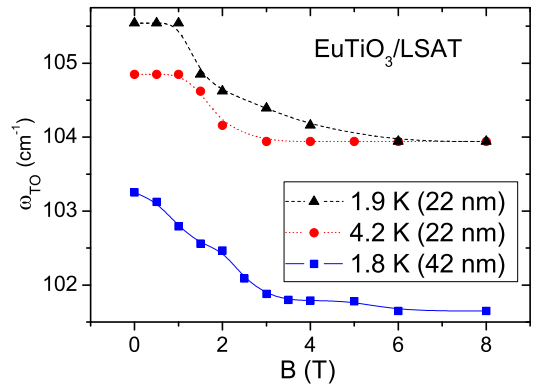}
  \end{center}
    \caption{(Color online) Magnetic field dependence of the lowest-frequency phonon obtained from the fits
    of the IR reflectance of both films at various temperatures below T$_{N}$.}
    \label{Fig10}
\end{figure}

\begin{figure}
  \begin{center}
    \includegraphics[width=70mm]{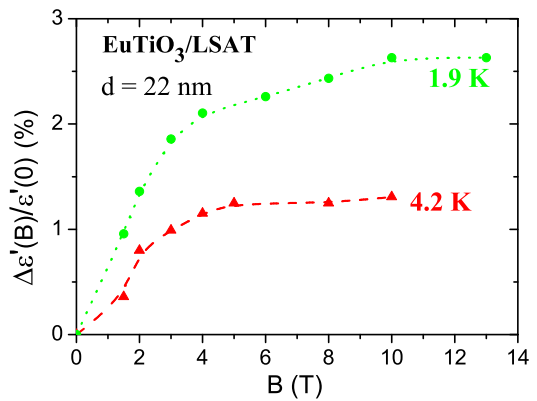}
  \end{center}
    \caption{(Color online) Magnetic field dependence of the relative changes of static permittivity obtained
    from the fits of IR reflectance of the 22 nm thin film.}
    \label{Fig11}
\end{figure}

Actually, we have observed an almost 2\cm\, decrease of the lowest-frequency phonon with
\textit{B} up to 10 Tesla (see Figs.~\ref{Fig9} and ~\ref{Fig10}). The phonon frequency
shifts could be determined with a high accuracy of 0.1\cm\, thanks to the sharp feature
of the reflectance band. The absolute changes of the phonon frequencies on \textit{B}
were similar and reproducible in both films, even though the origin of the sharp phonon
seen in Figs. 2 and 3 appears to be different. It corresponds predominantly to TO1 and an
optical phonon from the Brillouin zone edge (see the discussion in the next section) in
the 42 and 22 nm film, respectively. The higher frequency component of the TO1
reflectance band is weaker and broader, therefore it is not possible to evaluate its
change with \textit{B}. Magnetodielectric effect, i.e. the relative change of the static
permittivity with magnetic field \textit{B} ($\Delta\varepsilon'(B)/\varepsilon'(0)$)
obtained from the fits of reflectance spectra is plotted in Fig.~\ref{Fig11}. As
expected, one can see that the change of $\Delta\varepsilon'(B)/\varepsilon'(0)$ in the
thin film is almost three times smaller than that in the single
crystal,\cite{katsufuji01} because the phonons are stiffened in the films. The change of
$\Delta\varepsilon'(B)/\varepsilon'(0)$ appears to be larger at 1.9\K\, than at 4.2\K\,
(Fig.~\ref{Fig11}). It is caused by the larger phonon frequency shift with \textit{B} at
1.9\K\, (see Fig.~\ref{Fig10}), since also the shift of the phonon frequency on cooling
below T$_{N}$ is larger at 1.9\K\, than at 4.2\K\, (see inset of Fig.~\ref{Fig4}). It
corresponds also to larger relative change of the permittivity at 2\K\, than at 4.2\K\,
reported in single crystal\cite{katsufuji01} and ceramics.\cite{kamba07} Here it should
be noted that the magnetic field influence on the $A_{2u}$ SM component polarized
perpendicular to the film plane (unattainable in our experiment) should be higher than
the influence on the phonons polarized in the film plane. This is shown also in Table I.

Phonon frequencies obtained from the first principles calculation in the AFM (i.e.
without applied \textit{B}) and in the ferromagnetic phase (i.e. in a high \textit{B})
are summarized in Table~\ref{TableFreqs62}. One can see that mainly the lowest frequency
phonons shift down with the magnetic field. Theoretical frequency shift is twice larger
in Table~\ref{TableFreqs62} than in our experiment (Fig.~\ref{Fig10}), because the
calculations were performed at conditions of 0\K, while the experiment was done at
$\approx$2\K\, and the shift should increase on cooling below T$_{N}$. In this way we
confirmed that the magnetodielectric effect in EuTiO$_{3}$ is due to the change of the
polar phonon frequency with magnetic field.

It is well known from the literature that it is rather easy to
change the phonon frequencies by hydrostatic pressure or external
electric field. This effect is most remarkable in ferroelectrics,
in which the ferroelectric SM is highly sensitive to boundary
conditions. Electric field dependence of the ferroelectric SM was
first-time observed in Raman scattering of
SrTiO$_{3}$,\cite{schaufele67,worlock67} and the hydrostatic
pressure dependences of the SM were reviewed by Samara and
Peercy.\cite{samara81}

On the other hand, direct observations of phonon frequency changes on application of
external magnetic field \textit{B} are very scarce in literature and most of magnetic
materials do not exhibit measurable effects. Ruf et al.\cite{ruf88} observed 1.5\cm\,
change of one phonon line with \textit{B} in the Raman scattering spectra of
YBa$_{2}$Cu$_{3}$O$_{7}$. This phonon frequency was sensitive to the superconductive
phase transition and its changes were observed slightly below T$_{c}$ under the field of
6-12\,T. Sushkov et al.\cite{sushkov02} reported on small \textit{B}-dependent line-shape
changes in the IR absorption spectra of molecular magnet Mn$_{12}$-acetate, but a very
high magnetic field of 30\,T was necessary to get a measurable effect. Multiferroics
exhibit relatively high tunability of $\varepsilon$' so that also larger changes of
phonon frequencies with \textit{B} could be expected. This was confirmed in GdMnO$_{3}$
and DyMn$_{2}$O$_{5}$ multiferroics.\cite{pimenov06,cao08} In the former case, a phonon
frequency change by less than 1\cm\, was observed due to a phonon coupling with
magnon\cite{pimenov06} while in the latter case a high field of 18 T was necessary to see
a $\sim$2\cm\, change of phonon frequency.\cite{cao08} In our EuTiO$_{3}$ film the phonon
frequency change due to \textit{B} is comparable to the previous case, but a much lower
magnetic field was needed.

It should be stressed that the magnetodielectric effect in
EuTiO$_{3}$ is not caused by a linear magnetoelectric coupling,
since bilinear terms $\alpha_{ij}H_iE_j$ (where $H_i$ and $E_i$
are components of the magnetic and electric field, respectively)
are not allowed in the thermodynamic potential.\cite{shvartsman10}
Nevertheless, higher order magnetoelectric effects, accounted for
by the $\beta_{ijk}H_iH_jE_k$ and $\delta_{ijkl}H_iH_jE_kE_{l}$
terms in the thermodynamic potential, are permitted. The strong
quadratic and bi-quadratic magnetoelectric effects were actually
experimentally confirmed in bulk EuTiO$_{3}$ by Shvartsman et
al.\cite{shvartsman10}

Dzyaloshinskii recently proposed a splitting of polar phonons in ferroelectrics due to an
external magnetic field.\cite{dzyaloshinskii09} We do not see any splitting of phonons in
the magnetic field, but we observe a 50\% increase of the SM damping with magnetic field
(see the broadening of reflection band with magnetic field in Fig.~\ref{Fig9}). At
1.9\K\, its damping is 4.2\cm\, at \textit{B}=0 T, while at \textit{B}=13 T it is 6\cm.
Broadening of the phonon linewidth could be a consequence of a small phonon splitting,
which is not resolved.

\section{Discussion}

Let us now try to explain the origin of activation of the new mode near the TO1 phonon
frequency in the IR spectra of EuTiO$_{3}$/LSAT (see Fig. ~\ref{Fig4}). We consider two
possibilities: 1) the doubly degenerate $E_{u}$ component of the TO1 phonon probed in our
experiment splits due to anisotropic in-plane strain, which is created due to a phase
transition in the LSAT substrate. 2) the new mode originates from the Brillouin zone (BZ)
edge due to an AFD phase transition (with multiplication of the primitive unit cell and
BZ folding).

\begin{figure}
  \begin{center}
    \includegraphics[width=80mm]{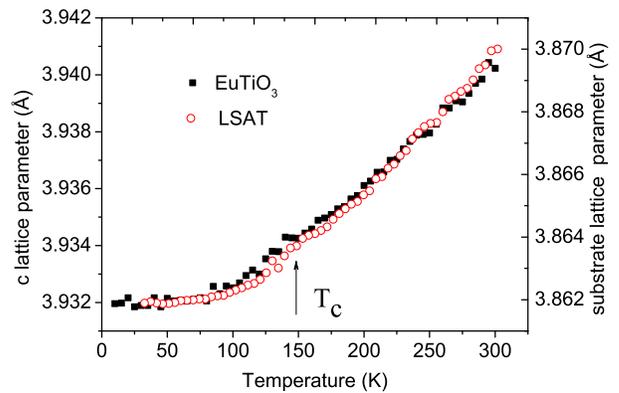}
  \end{center}
    \caption{(Color online) Temperature dependence of the out-of plane lattice
    parameter of the EuTiO$_{3}$ film (thickness 22\,nm) and of the LSAT substrate. }
    \label{Fig6}
\end{figure}

Let us discuss the first possibility. The in-plane polarized modes can split if the
tetragonal symmetry of film reduces to orthorhombic or lower-symmetry structure. It is
possible only in the case, when the in-plane strain would be anisotropic, i.e. the LSAT
substrate would not be cubic. Indeed, it has been reported that the LSAT substrate
exhibits a small distortion from cubic symmetry near 150\K\, and its low-temperature
phase has tetragonal or orthorhombic symmetry.\cite{chakoumakos98} We have measured
temperature dependence of the LSAT lattice constant and it really exhibits a small hump
near 150\K\, (see Fig.~\ref{Fig6}) and the same anomaly is revealed also in the out-of
plane EuTiO$_{3}$ film lattice parameter. Moreover, the structural phase transition in
LSAT can be recognized also in the microwave dielectric permittivity, where a small
anomaly was observed near 150\K\, (see Fig.~\ref{Fig7}). Nevertheless, in case of
substrate-induced anisotropic strain the split modes should have different symmetries and
therefore they cannot couple, as observed in our spectra. Moreover, the possible
anisotropic strain from the non-cubic LSAT substrate should as well split the phonons in
SrTiO$_{3}$/LSAT (SrTiO$_{3}$ and EuTiO$_{3}$ have the same lattice constant and
therefore the strain in both films is the same), but this was not
detected.\cite{nuzhnyy09} On the other hand, we have revealed the same new phonon
activated in IR spectra near 150\K\, in EuTiO$_{3}$ films deposited on NdGaO$_{3}$ and
LaAlO$_{3}$. Compressive strains in both films are slightly relaxed and have the value of
about 0.9\%,\cite{goian12a}, i.e. close to EuTiO$_{3}$/LSAT. Since the NdGaO$_{3}$ and
LaAlO$_{3}$ substrates do not exhibit any structural phase transition, the new mode
observed in all three films cannot be activated due to anisotropic strain.

Let us discuss the second possibility that the new mode originates
from the BZ edge. It can be expected that the AFD phase transition
evidenced in the bulk
EuTiO$_{3}$\cite{bussmann-holder11,allieta11,goian12a} and tensile
strained thin film EuTiO$_{3}$/DyScO$_{3}$ (see supplement of Ref.
\cite{lee10}) should occur in the compressive thin films as well.
The tetragonal $I4/mcm$ symmetry reported for the bulk can be also
well compatible with the thin film epitaxial geometry. Temperature
of the phase transition may be considerably shifted in comparison
to the bulk, of course. In the AFD phase two new $E_{\rm u}$
phonons near 255 and 430\cm\, stemming from the R-point of the BZ
edge become active in the IR spectra, as it can be seen from the
Table\,I. However, the frequencies are much higher than that of
our new mode seen between 110 and 105\cm.

It is known that two structural soft modes (SMs) from the R-point of the BZ edge can be
activated in low-frequency Raman spectra below the AFD phase transition in SrTiO$_{3}$
(T$_{C}$=105\K).\cite{fleury68} The AFD structure of SrTiO$_{3}$ is the same $I4/mcm$ as
in EuTiO$_{3}$ and the two AFD SMs should be IR inactive in both materials, if their
structure remains centrosymmetric. However, these AFD SMs were also discovered in the
low-temperature IR spectra of bulk SrTiO$_{3}$ ceramics\cite{petzelt01} and
polycrystalline films. \cite{ostapchuk02} Their activation was explained by polar
distortion near the grain boundaries.\cite{petzelt01,ostapchuk02} Activation of the AFD
SM was also observed in tensile strained SrTiO$_{3}$ and EuTiO$_{3}$ films on DyScO$_{3}$
substrates, since the tensile strain induces a ferroelectricity in both
films.\cite{lee10,nuzhnyy09b} Our new mode seen near 105\cm\, does not remind any AFD SM,
because it does not harden below the assumed AFD transition temperature near 150\K.
Nevetheless, it can be an optical mode from the BZ edge. Theoretical
calculations\cite{bettis11} show that the lowest-frequency optical branch exhibits almost
no \textit{q} dispersion in BZ, therefore TO1 mode (\textbf{q}=0) and the optical mode
from the BZ edge have almost the same frequency. AFD SMs have probably much lower
frequencies, therefore they are not resolved in our spectra. Nevertheless, both AFD SMs
and optical mode from BZ boundary can be activated after folding of BZ zone only if the
crystal structure becomes at least locally broken. Even if we know that the loss of the
inversion center was not evidenced by the SHG signal in EuTiO$_{3}$/LSAT,\cite{lee10}, it
is known that the IR probe is sensitive on local breaking of the symmetry known e.g. from
polar nanoclusters in relaxor ferroelectrics.\cite{hlinka06}

\begin{figure}
  \begin{center}
  \rotatebox{90}
    {\includegraphics[width=50mm]{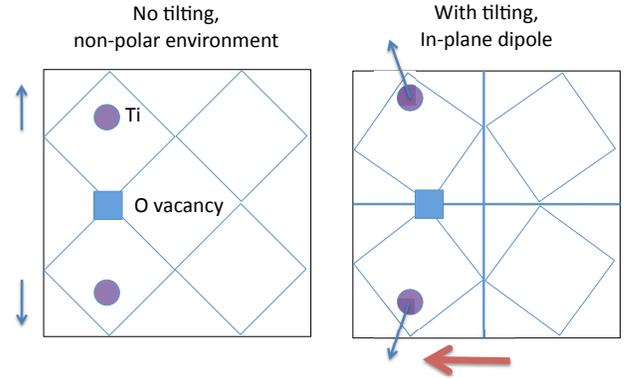}}
  \end{center}
    \caption{(Color online) Schematic plot of the perovskite crystal structure in the view along [001]
    axis in the a) untilted and b) tilted AFD phase. Squares are marks for the oxygen octahedra,
    Eu cations are not shown.
    Orientation of the electric dipole moment created by the Ti displacement in the vicinity of oxygen vacancy in the tilted
    phase is marked by a large red arrow.  }
    \label{O-vacancies}
\end{figure}

Here we propose a new possible mechanism of polar nanocluster appearance in the
EuTiO$_{3}$ as the consequence of AFD phase transition and oxygen vacancies, which are
always present in the lattice. Let us imagine that we have oxygen vacancies in the
TiO$_{2}$ plane of the perovskite structure (see Fig.~\ref{O-vacancies}). Then, in the
ideal perovskite structure without octahedra tilting the local environment around the
vacancy is antipolar: two neighboring Ti atoms are equally displaced in opposite
directions. In a structure which allows oxygen octahedra tilting, the displacements of Ti
atoms are antipolar in one direction, but with uncompensated dipoles in orthogonal
direction (see Fig.~\ref{O-vacancies}). In such case the local inversion symmetry is
broken. We have calculated the structure in a 40-atomic simulation cell (i.e.
EuTiO$_{2.875}$ instead of EuTiO$_3$) around this defect and found that it has caused Ti
displacements not only around the vacancy, but also in other Ti's in the vacancy
neighborhood. Of course, the displacements of the neighboring Ti are the largest. Also,
we found that the formation energy of the O vacancy in the TiO$_{2}$ plane is by 0.03 eV
lower than of the vacancy in the EuO plane (i.e. apical oxygen).

The IR inactive folded modes can become IR active also if we assume AFM short-range order
even if it remains centrosymmetric. Our density-functional calculations for an auxiliary
$I\overline{4}2m$ structure compatible with AFM order yielded a new IR active phonon
(originally optical one from the BZ edge) with frequency that is 12\cm\, lower than the
TO1 mode. The strength of this mode strongly depends on the conditions under which it is
calculated. We have used finite displacements approach to calculate the force constant.
For 0.025 $\AA$ displacement from their equilibrium position, the mode-plasma frequency
$\sqrt{\Delta\varepsilon}\omega_{TO}$ of the new mode is about 270\cm\, (for comparison
the TO1 mode has the plasma frequency around 1040\cm). If the displacement is only 0.0027
$\AA$, then the calculated mode-plasma frequency is only 24\cm. This is a sort of probe
of anharmonicity and it points to a rather strong anharmonicity of this particular mode.
This mechanism of the mode activation requires a short-range magnetic order at least up
to 150\K. Recent not yet published $\mu$SR experiment\cite{bosmann-holder12} performed on
bulk EuTiO$_{3}$ revealed short-range magnetic correlations up to 300\K, which is rather
surprising, because the AFM critical temperature is only 5.3\K.

Finally we can conclude that the new mode near 105\cm\, is of the same symmetry as the
TO1 mode. It probably stems from lowest-frequency optical phonon branch and it is
activated from the BZ edge due to the AFD phase transition. Its activation in IR spectra
requires local reduction of the symmetry, which can occur due to creation of polar or
magnetic nanoclusters.

\section{Conclusion}

Due to the compressive strain, all polar phonons polarized in the plane of the
EuTiO$_{3}$ films are considerably stiffened in comparison to the bulk ceramics. Below
150\K\, a new IR active mode was revealed near 110\cm. We understand it as optical phonon
from the BZ edge, which activates in the IR spectra below the AFD phase transition under
circumstance of presence of polar and/or magnetic nanoclusters. Magnetization data show
that the compressive strain enhances the N\'{e}el AFM temperature by 1\K\, due to the
strain enhanced superexchange interaction. IR reflectance studies in magnetic field
revealed a striking tuning of the lowest phonon frequency with the magnetic field. This
is responsible for the magnetodielectric effect reported previously in the bulk
EuTiO$_{3}$.\cite{katsufuji01} Observed change of the polar phonon frequency with the
magnetic field is in agreement with our theoretical calculations. Similar IR experiments
with magnetic field are rather rare in the literature, because the phonon frequency
changes are usually under spectral resolution in materials without a strong spin-phonon
coupling. Our study shows that the IR reflectance measurements of a rather thin film can
be more sensitive to the magnetic field (in the case of suitable non-conducting
substrate) than the reflectivity spectra of the bulk samples or transmission spectra of
the thin films. So, the IR reflectance spectroscopy is a very promising tool for a study
of the magnetodielectric effect as well as of structural phase transitions in thin
dielectric films deposited on electrode-less nonconducting substrates.

\begin{acknowledgments}

This work was supported by the Czech Science Foundation (Projects Nos. 202/09/0682 and
P204/12/1163). J.H. Lee and D. G. Schlom were supported by the National Science
Foundation through the MRSEC program (Grant No. DMR-1120296). T. Birol and C. J. Fennie
were supported by the DOE-BES under Grant No. DE-SCOO02334. Part of this work was
supported by the Young Investigators Group Programme of the Helmholtz Association,
Germany, contract VH-NG-409. KZR and ML gratefuly acknowledge the support of J\"{u}lich
Supercomputing Centre. We are grateful to E. \v{S}antav\'{a} for help with the magnetic
measurements.

\end{acknowledgments}

\end{document}